# Resistance to Critiques in the Academic Literature: An Example from Physics Education Research

CHARLES REICHHARDT[*], ALEX SMALL[**], CRISTIANO NISOLI[*], & CYNTHIA REICHHARDT[*]

[*]Los Alamos National Laboratory, Los Alamos, New Mexico, USA. Email cjrx@lanl.gov
[**]California State Polytechnic University, Pomona, California, USA

Research framed around issues of diversity and representation in STEM is often controversial. The question of what constitutes a valid critique of such research, or the appropriate manner of airing such a critique, thus has a heavy ideological and political subtext. Here we outline an attempt to comment on a paper recently published in the research journal *Physical Review—Physics Education Research* (PRPER). The article in question claimed to find evidence of 'whiteness' in introductory physics from analysis of a six-minute video. We argue that even if one accepts the rather tenuous proposition that 'whiteness' is sufficiently well defined to observe, the study lacks the proper controls, checks, and methodology to allow for confirmation or disconfirmation of the authors' interpretation of the data. The authors of the whiteness study, however, make the stunning claim that their study cannot be judged by standards common in science. We summarize our written critique and its fate, along with a brief description of its genesis as a response to an article in which senior officers of the American Physical Society (which publishes PRPER) explained that the appropriate venue for addressing issues with the paper at hand is via normal editorial processes.

**Introduction**

Productive critique and debate are essential elements of most truth-seeking enterprises. While political and ideological biases pose risks for any area of scientific debate, those risks are heightened in areas such as science education research, where values-laden controversies from the wider society intertwine with more practical, empirical questions about educational methods and measured performance. Open, transparent debate is thus of especial significance when the question facing scientists is not 'What could these data tell us about the natural world?', but rather 'What are the obstacles to teaching people how to interpret data about the natural world?'

In recent years, education journals in physics and chemistry have published a number of articles that approach educational issues from very explicitly values-laden, justice-oriented perspectives, e.g.*,* on matters of race (Robertson and Hairston 2022; Robertson *et al.* 2023), colonialism (Reyes *et al.* 2023), and disability (Robertson 2023). There is a commendable aspect to scholars making their ideological commitments explicit, if it helps the scholarly community calibrate for potential biases. Indeed, candid disclosure of biases and perspectives can be a first step towards adversarial collaborations, between researchers with opposing beliefs or goals, an increasingly recognized practice in social science (Nosek and Errington 2020). However, when a work is explicitly framed in terms of politically ascendant ideologies, there is a risk that potential critics may be intimidated at the prospect of openly disagreeing. It is thus essential that scientific publishers pay extra attention to such work and be even more fastidious about allowing room for productive debate and discussion.

Here we discuss an example of a prominent journal impeding reasoned critique of a controversial article involving matters of race and equity in science education. In March 2022, the journal *Physical Review—Physics Education Research* (PRPER) published an article titled 'Observing whiteness in introductory physics: A case study' (Robertson and Hairston 2022). Robertson and Hairston claimed that their case study of three students solving a problem with some guidance from an instructor indicated the presence of 'whiteness' in the physics classroom. While there were several elements to the concept of 'whiteness' as used by Robertson and Hairston, the aspect most specifically relevant to the classroom interactions that they analyzed was that whiteness requires a centre. Robertson and Hairston saw the concept of a centre manifesting every time the instructor attempted to focus students' efforts on a specific diagram. They offered no means by which one could falsify their broader conclusions about the physics community.

The paper elicited controversy, unsurprisingly, leading to editorials in both PRPER (Henderson and Thoennessen 2022) and *APS News* (Hellman *et al.* 2022), an official newsletter of the American Physical Society (APS, publisher of PRPER). Both editorials quite appropriately denounced the (sadly unsurprising) harassment and hate mail that Robertson and Hairston experienced in the wake of their article. Both editorials also made clear that reasoned, respectful commentary is welcome. The Lead Editor and Editor in Chief of PRPER wrote 'The *Physical Review* invites constructive and respectful criticism of published articles in the form of Comments' (Henderson and Thoennessen 2022). This call was echoed in *APS News*, where an editorial by senior leaders of APS quoted the same sentence from PRPER and added:

> We encourage the physics community to approach physics education research as they would any field outside their immediate area of expertise: with curiosity, skepticism, and an open mind. Progress in research requires critical examination; ideas and results are often challenged; immediate consensus is not expected. Constructive criticism is a way to support researchers and promote their efforts, but harassment and personal attacks have no place in scientific discourse. (Hellman *et al.* 2022)

In response to that call, we prepared a comment on the paper 'Observing whiteness' and submitted it via standard journal submission processes. Below, we summarize the article and its shortcomings, and then recount the response of reviewers and editors when we submitted our critique to the journal as a comment via standard editorial processes. Most of the text that follows is verbatim from our original comment to the journal (which we have since published on a preprint server; Reichhardt *et al.* 2023), so that the reader can judge the substance and tone of our critique. Separate from the question of whether our arguments ultimately outweigh those of the authors (and we certainly believe that they do) is the question of whether our arguments were presented in a manner appropriate to scholarly dialogue. If, as we believe, our comment was reasoned and appropriate, then its rejection by the journal should be concerning to anyone who wants to see a thriving, constructive body of literature in the field of science education research.

## Critique of the Whiteness Article

Robertson and Hairston analyze a six-minute video of three students and an instructor discussing physics problems. The comments, actions, and interview responses of the participants are analyzed and discussed under the lens of Critical Race Theory and Critical Whiteness Studies. They made much of the instructor's attempts to focus students' attention on a diagram, and equated this with the notion of whiteness requiring a centre. We argue that even if one accepts the questionable proposition that this constitutes an accurate (or even coherent) notion of 'whiteness', the study does not have proper controls or checks, and lacks a methodology that would allow other follow-up workers to confirm or disconfirm the asserted interpretation of the data, or even to repeat the study. It is plausible that other observers might see multiple layers to the instructor's actions, rather than an overriding, constraining focus.

Unfortunately, the authors attempt to foreclose the possibility of critiques from the physics community, arguing that 'physics as an epistemology is not well-suited to assess arguments of the kind we have made here' (Robertson and Hairston 2022: 16) and dismiss questions like 'But is that really whiteness?' as 'bad faith' (Robertson and Hairston 2022: 17). Their framework is thus presented as completely incompatible with quantitative and qualitative scientific studies and has no place in the scientific journals of *Physical Review*.

One key issue is that the study lacks sufficient data to draw any broad conclusions about physics as a community. Only a single video interaction is discussed and there is no broader background data provided against which conclusions could be drawn. It can be argued that this is a qualitative study intended merely to illustrate concepts that can be further explored in subsequent work; however, qualitative studies are most illuminating when multiple interpretations are discussed and weighed, along with other possible validity threats to the conclusions (Ragin *et al.* 2004).

This is particularly important since only one video was considered, and it is not clear from the study whether the interactions are representative of broader introductory physics interactions. For example, if a video were recorded of a different set of students, entirely different approaches or comments could appear. Ideally, it would be better to analyze an ensemble of such videos to see if repeated themes occur, and also to test whether there is consistency in the analysis. Even a purely methodological paper, intended only to illustrate a method rather than use it to draw firm conclusions, is most useful when the methodology is shown to be robust.

The authors do not provide evidence for the reliability of their data. The small amount of video data presented appears to be of poor quality since the transcription of the video is riddled with unintelligible remarks. Also, in addition to offering no mechanism for mitigating possible self-selection bias among the students interviewed (Shubert and Meredith 2015) or for avoiding judgment biases in the simulated recall interviews (Lyle 2003), the study lacks proper controls of the viewers of the interactions that would make it possible to conclude whether they generate consistent definitions of the same actions, thereby giving the reader confidence that similar results would be obtained by an independent researcher employing the same methodology (Shubert and Meredith 2015).

For example, if an ensemble of videos were analyzed, would a consistent analysis appear in which one group of students presenting a behavior opposite to that of another group of students is detected as a clear difference? In this article by Robertson and Hairston, there is no methodology that would ensure that different people viewing the video for the purpose of observing whiteness would give the same analysis for the interactions, comments, or actions of any student or the instructor. One method of providing such control is the inter-rater reliability method employed in previous physics education research (PER) studies (Scherr 2009).

Another method would be to consider an ensemble of videos, which would permit consistency checks to be established and some type of statistical analysis to be performed. For example, if 100 videos were analyzed, it could be determined what fraction indicates witnessing whiteness compared with what fraction does not. Additionally, the degree to which whiteness is observable in individual videos could be measured, as well as how many times whiteness was completely absent. From a single brief video it is not possible to determine whether certain types of introductory-physics problems allow for easier observations of whiteness than other problems. There is no basis on which to conclude that any and all problem-solving activities would lead reasonable observers to see similarly problematic examples of whiteness in the classroom.

Furthermore, the study lacks a mechanism for disconfirmation, a key aspect of any scientific work (Popper 1959; Maxwell 2004). Robertson and Hairston do not give significant discussion to whether alternative interpretations of the classroom situation might be plausible or reasonable. What actions or statements of a student or instructor would indicate that whiteness is not manifesting in the interaction? Any group trying to reproduce this work will likely observe different interactions from the single video shown in this study, so some form of framework that would indicate whether new observations are consistent or inconsistent with the conclusions of the present study is needed but is not provided.

One useful method would be a table of interactions or statements that are indicative of whiteness, along with a similar table listing interactions or statements that indicate the absence of whiteness, as well as a list of interactions or statements from which no interpretation can be drawn. Such a table would also help ensure that the study avoids confirmation bias (Nickerson 1998; Maxwell 2004).

In addition, the study could include videos in which participants act or speak in the opposite manner during the interactions, and have multiple observers perform an analysis to see whether they draw the same or different conclusions from the video. This would provide confidence in the validity of the analysis.

Many of our suggestions involve making the study more quantitative. This is a natural direction for a journal in the *Physical Review* family, since physics is a quantitative science. We respect that the social sciences often employ qualitative methods in order to develop new theories that can later be subjected to quantitative research. It is important to keep in mind, however, that social science is still a science, and as such, its practitioners construct models of the world based on evidence (Russ and Odden 2017) and perform tests of these models through transparent, well documented procedures.

Qualitative research studies focused on case studies can, of course, be appropriate in science journals if used to highlight questions and hypotheses worthy of further examination, much as theoretical-physics articles often examine deliberately unrealistic 'toy' models to stimulate further inquiry. Even as we ultimately draw conclusions about the physical world not from theoretical analysis of simplistic toy models but from comparing careful theoretical models with experiments on real systems, large-scale conclusions about problems in the wider physics community require quantitative research on large samples with proper validation.

We finally wish to comment on the general concept of testing competing hypotheses and why it is important to have statistical significance measures and consistent interpretations of observations in studies of this type. In their opening sentence Robertson and Hairston write, 'Critical Race Theory names that racism and white supremacy are endemic to all aspects of U.S.

society, from employment to schooling to the law', and in discussing their results, they write, 'That whiteness is "ordinary" in physics classrooms is not surprising, given critical race theory's assertion that whiteness is endemic to every aspect of U.S. society.' Under this conjecture, observation of any phenomenon will, by definition, be consistent with whiteness since whiteness is 'endemic' in everything. This is a circular argument in which the hypothesis is defined to be true for every possible observation, giving it no scientific validity. To qualify as scientific, a study must enable development of predicted observations that would nullify or modify the conjecture.

In other words, it must be possible to disconfirm as well as confirm the hypothesis (Platt 1964), a tenet that physics education research works to help instill in students (Russ and Odden 2017).

Under repeated observations, there would likely be a range of outcomes and even opposite outcomes, and the results should be discussed in the light of possible competing interpretations. For example, a student interaction scenario in Doucette and Singh (2021) that is very similar to the interactions considered by Robertson and Hairston led to a very different interpretation in terms of female students assuming leadership roles. It would be valuable to identify methodologies that could distinguish between the interpretations of Robertson and Hairston and those of Doucette and Singh (2021). Alternatively, some modification of the theory might need to be made. If the paper's hypothesis is scientific in the sense accepted by other physics journals, it should be susceptible to disconfirmation as well as confirmation.

### Reception of the Critique

'Observing whiteness in introductory physics: A case study' (Robertson and Hairston 2022) was published in the American Physical Society (APS) Open Access journal *Physical Review – Physics Education Research* (PRPER), in March 2022. During the following month, accounts of the PRPER paper appeared in the popular press and in social media, generating controversy. In response, in April 2022, PRPER editors published 'Editorial: Research on Advancing Equity is Critical for Physics' (Henderson and Thoennessen 2022) in which the Editors wrote, 'The Physical Review invites constructive and respectful criticism of published articles in the form of Comments.'

Shortly afterwards, in May 2022, *APS News* published the article 'Productive Scientific Discourse Demands Respect' (Hellman et al. 2022). We contacted the Back Page authors, challenging their implication that disrespectful communications are originating from APS members, and expressing doubt that a Comment can, as suggested in the PRPER editorial and *APS News* article, be used as a venue for discussion of the paper. In June 2022, then-APS President Hellman responded and encouraged us to write and submit 'a comment based on a scientific response, presumably framed in a constructive and respectful way, addressing the topics you found problematic' in the PRPER article. We complied and submitted our Comment to PRPER.

Robertson and Hairston declined to respond to our comment, and after the PRPER Editor received two referee reports, in September 2022 he rejected our Comment on the grounds that it was 'framed from the perspective of a research paradigm that is different from the one of the research being critiqued'. We contacted Hellman *et al.* a second time, stating that although the PRPER Editorial, the Back Page article, and the APS President indicated that the appropriate venue for discussion is a Comment, PRPER was not making that venue available. The APS

President recommended that we appeal the rejection of our Comment, so we submitted a formal appeal in October 2022. In January 2023, the PRPER Editor rejected our appeal on the grounds that one referee and an Editorial Board member upheld the original rejection that our Comment was not written from the same paradigm used by Robertson and Hairston in their work.

We filed a formal appeal with the Editor-in-Chief of *Physical Review* in February 2023. In March 2023, the Editor-in-Chief rejected our appeal, stating that

> the role of Comments in the *Physical Review* is to refute or correct specific results in the articles upon which they are focused. In this case, your Comment was a broader commentary on the nature of all studies conducted like those in the original article. For that reason, I do not believe that your concerns should appear in this context.

After receiving this report, we made our Comment publicly available (Reichhardt *et al.* 2023) to clarify to the community that *Physical Review* policy contradicts the assertions in the PRPER Editorial and the *APS News* article that a Comment is the appropriate venue for a discussion of the paper.

**Conclusions**

It is disturbing that a prestigious scientific journal has not made good on its invitation for comments on a controversial article. Our critique is, of course, not infallible, and certainly there are defenses that could be made for the whiteness article by Robertson and Hairston. For instance, careful analysis of a case study can illustrate key elements of a theoretical framework, and thus qualitative analysis of case studies deserves leeway in recognition of that role. However, the closing section of their paper makes clear that the goal is not merely to illustrate methodology, but to illustrate points about the wider world, and we stand by our critique that conclusions about the wider world require a wider data set. Reasonable researchers might, of course, disagree, and those disagreements should be aired through open, respectful debate. Otherwise, it is difficult for scholarly communities to pursue their truth-seeking task.

It would be tempting to ascribe the rejection of our comment to specific defects in our reasoning or writing. However, there is a pattern of resistance to critique at the journal in question. A search of the *Physical Review* archives indicates that since its inception in 2005, PRPER has only published one comment article, despite publishing over 1000 articles in total (Rieger *et al.* 2016). During that same time period, *Physical Review Letters* (PRL) (widely considered the most prestigious and selective journal published by APS) published 792 comments on 58,000 articles. On a comments-per-article basis, PRL is more than 10 times as open to critique as PRPER.

There is no reason to believe that the paucity of comments is due to fewer errors in PRPER articles. It is similarly unlikely that physicists are so uniform in their viewpoints on education that essentially everyone in the community agrees with the perspectives offered in the pages of PRPER, and are hence unable to formulate dissenting thoughts. Rather, it seems that this journal, notionally approving of diversity, is inhospitable to diversity of opinion, and the dissent that inevitably follows. Although Robertson and Hairston presented their Critical Race Theory (CRT) framework as if it falls within a scientific approach, they also rejected the possibility of good-faith critique of their work. When their study was nonetheless challenged within a scientific framework, the editors and reviewers apparently adopted the stance of Robertson and Hairston, arguing that such challenges should not be allowed because it is necessary first to accept the authors' framework prior to offering a critique. This is akin to stating that an astronomer must first accept astrology as true before critiquing it. Such notions should be, at a minimum, dispiriting for anyone who sees educational practices as worthy of empirical investigation.

**Competing Interests**

The authors declare that there are no competing interests.

**About the Authors**

**Charles Reichhardt** is a physicist conducting research in the Theoretical Division at Los Alamos National Laboratory.

**Alex Small** is a full professor of physics (with tenure) and department chair at California State Polytechnic University Pomona.

**Cristiano Nisoli** is a physicist conducting research in the Theoretical Division at Los Alamos National Laboratory.

**Cynthia Reichhardt** is a physicist conducting research in the Theoretical Division at Los Alamos National Laboratory.